\documentclass[journal,twoside,web]{ieeecolor}
\usepackage{generic}
\usepackage{cite}
\usepackage{amsmath,amssymb,amsfonts}
\usepackage{algorithmic}
\usepackage{graphicx}
\usepackage{textcomp}
\usepackage{amsmath}
\usepackage{algorithm,algorithmic}
\usepackage{hyperref}
\usepackage{graphicx}
\usepackage{amsmath}
\usepackage{booktabs}
\usepackage{caption}
\usepackage{geometry}
\usepackage[T1]{fontenc}
\hypersetup{hidelinks=true}
\usepackage{textcomp}
\def\BibTeX{{\rm B\kern-.05em{\sc i\kern-.025em b}\kern-.08em
    T\kern-.1667em\lower.7ex\hbox{E}\kern-.125emX}}
\begin{document}
\title{CDI-DTI: A Strong Cross-domain Interpretable Drug-Target Interaction Prediction Framework Based on Multi-Strategy Fusion}
\author{Xiangyu Li, Haojie Yang, Kaimiao Hu, Runzhi Wu, Liangliang Liu and Ran Su
\thanks{This study was supported by the National Natural Science Foundation of China (Grant No. 62222311); corresponding author: Liangliang Liu and Ran Su}
\thanks{Xiangyu Li, Haojie Yang, Kaimiao Hu , Runzhi Wu and Ran Su are with Department of Intelligence and Computing, Tianjin University, Tianjin 300350, China (e-mail: {xiangyuli,haojae,kaimiaohu,wurunzhi,ran.su}@tju.edu.cn)}
\thanks{Liangliang Liu is with the College of Information and Management Science, Henan Agricultural University, Zhengzhou 450002, China (e-mail: liangliu@henau.edu.cn)}}
\maketitle
\begin{abstract}
Accurate prediction of drug-target interactions (DTI) is pivotal for drug discovery, yet existing methods often fail to address challenges like cross-domain generalization, cold-start prediction, and interpretability. In this work, we propose CDI-DTI, a novel cross-domain interpretable framework for DTI prediction, designed to overcome these limitations. By integrating multi-modal features-textual, structural, and functional-through a multi-strategy fusion approach, CDI-DTI ensures robust performance across different domains and in cold-start scenarios. A multi-source cross-attention mechanism is introduced to align and fuse features early, while a bidirectional cross-attention layer captures fine-grained intra-modal drug-target interactions. To enhance model interpretability, we incorporate Gram Loss for feature alignment and a deep orthogonal fusion module to eliminate redundancy. Experimental results on several benchmark datasets demonstrate that CDI-DTI significantly outperforms existing methods, particularly in cross-domain and cold-start tasks, while maintaining high interpretability for practical applications in drug-target interaction prediction.
\end{abstract}

\section{Introduction}
Accurate prediction of drug-target interactions (DTI) has become a crucial component in the modern drug discovery and development pipeline \cite{chen2016drug, wen2017deep, chen2018machine}. Although traditional biochemical experiments play an irreplaceable role in drug discovery, large-scale identification of novel DTIs faces numerous challenges, often consuming enormous human, material, and financial resources and taking a considerable amount of time \cite{dimasi2003price, paul2010improve, pushpakom2019drug}. Therefore, developing efficient and accurate computational methods for DTI prediction holds significant strategic value \cite{zhang2022deepmgt}.

Recently, machine learning and deep learning-based methods have emerged as the mainstream in DTI prediction \cite{bagherian2021machine, ding2014similarity, wang2023fusion, nguyen2021graphdta}. These models explore the intrinsic characteristics of drugs and proteins and integrate various neural network components to construct predictive frameworks. For example, Öztürk et al. first proposed DeepDTA, which employs convolutional neural networks (CNNs) and multilayer perceptrons (MLPs) to extract molecular and protein features for DTA estimation \cite{ozturk2018deepdta}. Building upon this foundation, many studies have improved model performance by enhancing feature extraction \cite{lee2019deepconv, tsubaki2019compound}. Lee et al. \cite{lee2019deepconv} introduced DeepConv-DTI, leveraging one-dimensional CNNs (1D-CNNs) to extract features from amino acid sequences. Huang et al. \cite{huang2021moltrans} proposed MolTrans, where Transformer-encoded drug SMILES and protein sequence embeddings are mapped into a two-dimensional binding map, subsequently processed by 2D-CNN and fully connected layers for interaction prediction.

Other studies have focused on optimizing the classifier design to improve DTI prediction. For instance, Zhao et al. \cite{zhao2022hyperattentiondti} introduced HyperAttentionDTI, incorporating a HyperAttention mechanism to enhance feature representation, while other works designed multi-head cross-attention modules to further strengthen interaction modeling \cite{zhao2022hyperattentiondti, zeng2021deep, bian2023mcanet}. Furthermore, numerous graph neural network (GNN)-based methods have been proposed for DTI prediction. In these frameworks, drugs and proteins are represented as graphs encoding atomic and amino acid relationships, from which graph features are extracted for prediction \cite{tsubaki2019compound, cheng2022iifdti}. With the rapid development of large language models (LLMs), pretrained embeddings have also been widely adopted in DTI prediction. For instance, Lee et al. \cite{lee2024dlm} proposed DLM-DTI, which employs a teacher-student architecture to distill a lightweight version of ProtBERT for protein embeddings and combines them with ChemBERTa-derived drug embeddings for MLP-based prediction. Despite the progress of these approaches on standard single-domain datasets, their generalization ability remains limited in real-world scenarios such as identifying novel drugs or targets for complex diseases, showing behavior similar to the “cold-start” problem in recommender systems. This limitation primarily stems from the restricted representational capacity of current models: most rely heavily on prior knowledge from source-domain training data, such as pretrained embeddings or graph-based representations ( Word2Vec \cite{chen2020transformercpi}, Smi2Vec \cite{lin2020deepgs}, and GNN-based molecular graphs \cite{ye2022molecular}). Consequently, these models struggle to generalize to unseen structures or novel semantic patterns, leading to significant performance degradation under cross-domain or cold-start conditions.

With the advent of bioinformatics and the increasing availability of heterogeneous biomedical data, Multi-modal DTI prediction has emerged as a promising direction to enhance representational diversity and robustness \cite{zhou2021multidti}. For instance, MFCM-DTI \cite{li2024mfcm} integrates molecular structure embeddings, bioactivity embeddings, and protein sequence features to capture the complex interactions between drug molecules and key amino acids of target proteins. Hu et al. \cite{hu2025multi} leveraged the UnitedDTA framework to combine 1D, 2D, and 3D representations of drugs and proteins via contrastive learning and cross-attention mechanisms, achieving unified discriminative embeddings across modalities. Similarly, MMDG-DTI \cite{hua2025mmdg} fused textual and structural representations of drugs and proteins using domain-adversarial training (DAT) and contrastive learning to enhance generalization capability. However, Du et al. \cite{du2022compound} argued that existing Multi-modal methods still fail to fully exploit cross-modal information. A major challenge lies in designing effective fusion strategies. Early fusion methods first integrate multiple modalities of a drug or protein to construct unified representations before performing DTI modeling. This allows a comprehensive and semantically consistent description of molecular entities. However, early fusion often overlooks cross-modal interactions between drugs and targets, limiting fine-grained correspondence learning. Conversely, late fusion models conduct interaction modeling within each modality and subsequently combine results from different modalities. While this approach better captures complementary cross-modal information, it is prone to redundancy and noise accumulation during the final integration stage, undermining robustness and discriminative power. These challenges highlight the urgent need for a balanced Multi-modal fusion strategy for DTI prediction.

To this end, we propose CDI-DTI, A Strong Cross-domain Interpretable Drug-Target Interaction Prediction Framework Based on Multi-Strategy Fusion. CDI-DTI aims to enhance DTI prediction accuracy through staged Multi-modal feature fusion. It captures textual, structural, and functional characteristics of drugs and proteins. Specifically, we employ ChemBERTa and ProtBERT to extract contextual embeddings from SMILES strings and amino acid sequences, respectively. Structural features are derived from SMILES-based molecular graphs and AlphaFold-predicted protein structure graphs, capturing topology-level information. Functional descriptions are obtained via the generative model MoT5 and the DeepGO annotation database, with embeddings encoded using BioBERT.

During training, CDI-DTI performs multi-stage fusion of these three modalities to ensure efficient representation alignment. Extensive experiments on multiple benchmark DTI datasets validate the effectiveness of our framework. The main contributions of this work are summarized as follows:
\begin{itemize}
\item[$\bullet$] We propose CDI-DTI, a novel Multi-modal multi-stage fusion multi-task learning framework, which constructs textual, structural, and functional features from standard DTI datasets using deep learning methods, thereby improving prediction accuracy and generalization.
\item[$\bullet$] We introduce a balanced fusion strategy that effectively combines early and late fusion techniques. A multi-source cross-attention module is designed, and Gram Loss is incorporated to align and fuse multi-modal data of drugs and targets. Additionally, a deep orthogonal fusion module is introduced to mitigate feature redundancy in the late fusion process. enabling deep semantic representation of drug-target interactions.
\item[$\bullet$] Extensive experiments on multiple benchmark datasets demonstrate the robustness of CDI-DTI, showcasing significant improvements in cold-start and cross-domain scenarios.  The framework excels in generalization across different domains and achieves state-of-the-art performance in predicting novel drug-target interactions. 
\item[$\bullet$] We validate its interpretability, providing valuable insights for precision drug design by visualizing the learned feature interactions and offering clear explanations for the model's decision-making process.
\end{itemize}
In addition, a preliminary version of this work has been reported\cite{li2025m3st}.
\section{Materials and Methods}
\subsection{Benchmark Datasets}
In this study, we employ the BindingDB and DAVIS datasets to train and evaluate model performance. The BindingDB dataset \cite{liu2007bindingdb} consists of 10,665 drugs and 1,413 proteins, with 32,601 measured interactions based on dissociation constant (Kd). We defined drug-protein interactions as positive samples and unobserved interactions as negative samples, with 9,166 positive and 23,435 negative samples, respectively. The DAVIS dataset \cite{davis2011comprehensive} contains 68 drugs and 379 proteins, with 11,103 measured interactions in terms of Kd. Interactions are also divided into positive and negative classes, with 1,506 and 9,597 samples, respectively. Details of both datasets are summarized in Table~\ref{table 1}.
\begin{figure*}
\centerline{\includegraphics[width=\textwidth]{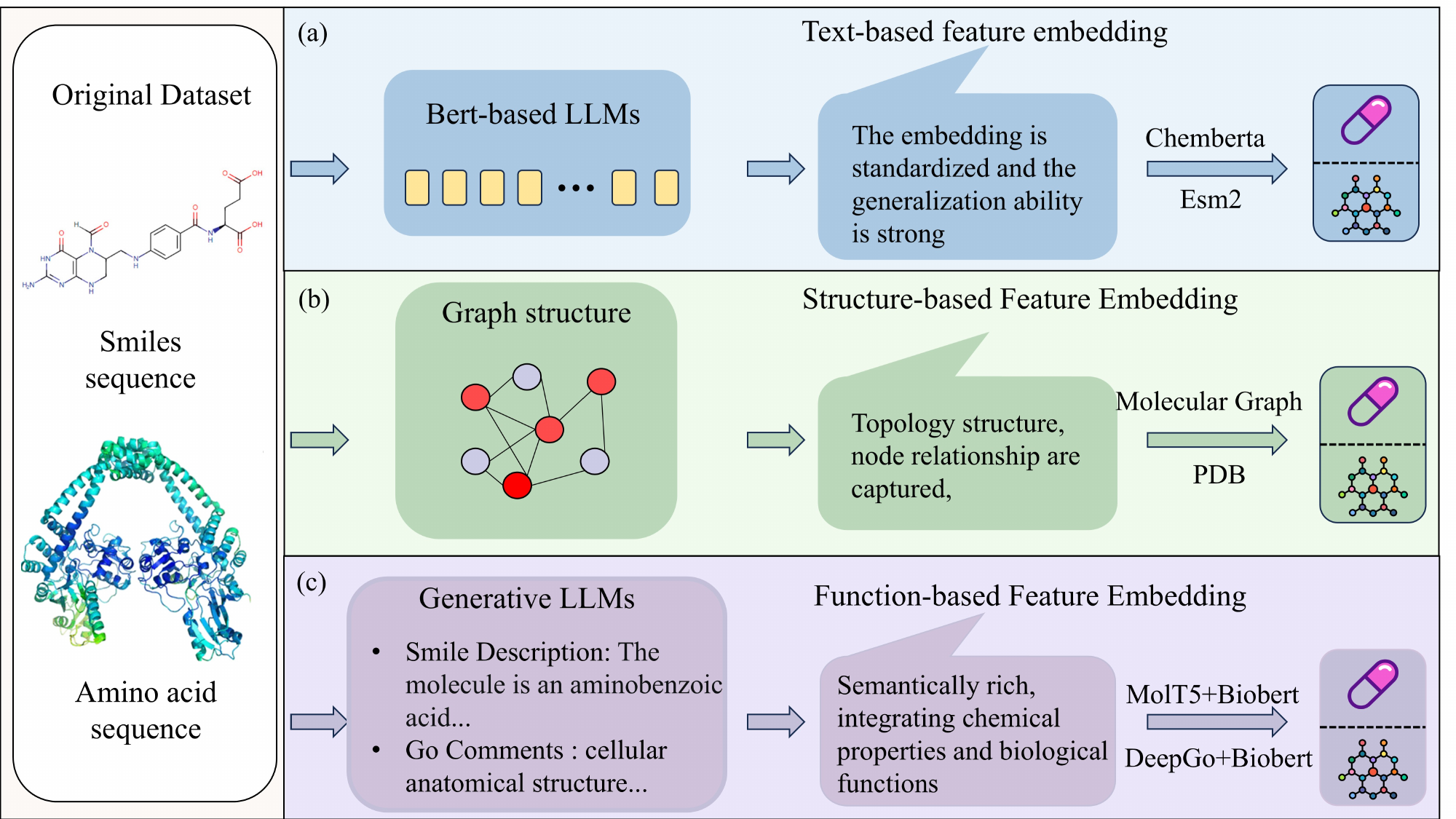}}
\caption{Multi-modal Feature Construction. (a) Textual features are extracted from drug SMILES and protein sequences using ChemBERTa and ESM2, respectively. (b) Structural features are derived from drug molecular graphs constructed via RDKit and protein structure graphs obtained from AlphaFold. (c) Functional features are generated from drug descriptions produced by MolT5 and protein GO annotations predicted by DeepGO, both encoded using BioBERT.}
\label{fig 1}
\end{figure*}
\subsection{Multi-modal Feature Construction}
In the CDI-DTI framework proposed in this study, we comprehensively represent protein and drug molecular features from three embeddings: textual, structural, and functional embeddings. All modal features are derived from sequence information and extracted through mainstream pretrained models to ensure diversity in feature representation and biological interpretability\cite{chithrananda2020chemberta ,lin2022language,jumper2021highly,kulmanov2018deepgo,lee2020biobert,edwards2022translation}. Therefore, we first complete the construction of the three modal features. The process is illustrated in Figure~\ref{fig 1}, and the detailed process is shown in appendix.

Through the above processing, we obtain high-quality Multi-modal features encompassing semantic $x_{t}$, structural $x_{g}$, and functional $x_{f}$ dimensions. Ultimately, we collected 32,601 samples from the BindingDB dataset and 11,103 samples from the DAVIS dataset. To ensure fair comparisons of model performance, we used the same training, validation, and test datasets as in previous studies \cite{huang2021moltrans, kang2022fine}, as shown in Table~\ref{table 2}.
\begin{table}
\caption{Statistics of Benchmark Datasets}
\label{table 1}
\begin{tabular}{|c|c|c|c|}
\hline
Dataset&Drug Samples&Protein Samples&Interactions\\
\hline
BindingDB&10,665&1,413&32,601\\
\hline
DAVIS&68&379&11,103\\
\hline
\end{tabular}
\end{table}
\begin{table}
\caption{Data Splits of Benchmark Datasets}
\label{table 2}
\begin{tabular}{|c|c|c|c|}
\hline
Dataset&Training Set&Validation Set&Test Set\\
\hline
BindingDB&12,668&6,644&13,289\\
\hline
DAVIS&2,086&3,006&6,011\\
\hline
\end{tabular}
\end{table}
\subsection{CDI-DTI Method}
We propose a Multi-modal and multi-stage feature fusion multi-task learning framework (CDI-DTI), aiming to systematically improve DTI prediction accuracy through Multi-modal data integration. As illustrated in Fig. 2, the CDI-DTI framework consists of four major stages: Multi-modal feature extraction (Fig. 2(a)), early fusion (Fig. 2(b)), late fusion (Fig. 2(c)), and classification head.
\subsubsection{Multi-modal Feature Extraction Stage}
The goal of the Multi-modal feature extraction stage is to refine the representations of textual, structural, and functional modalities from the embeddings of drug molecules and proteins, thereby providing rich information representations for subsequent fusion and prediction.
\begin{figure*}[t!]
\centering
\includegraphics[width=\textwidth]{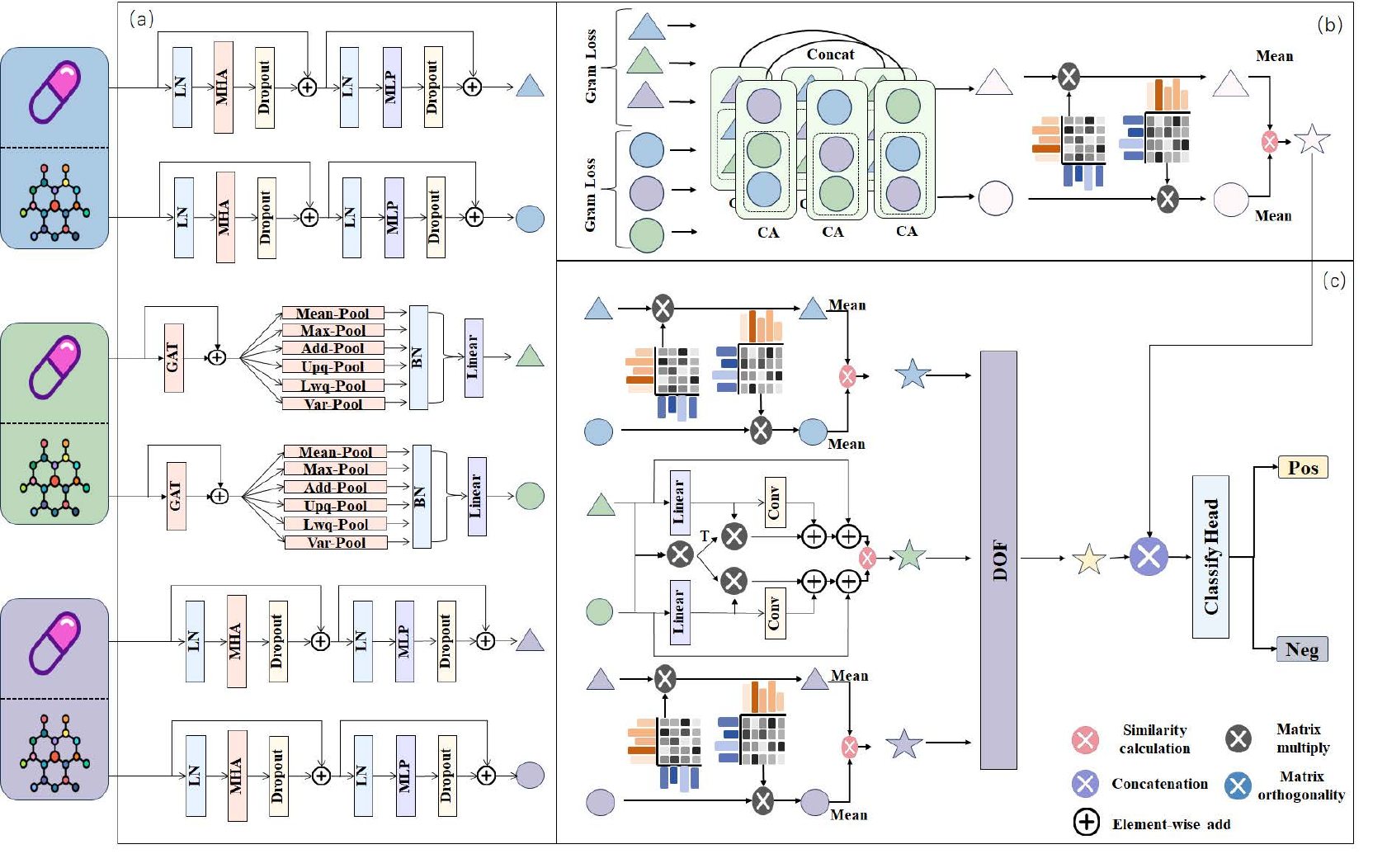} 
\caption{Overview of the  CDI-DTI Framework. (a) Multi-modal feature extraction stage: Textual features are refined using a multi-head self-attention module; structural features are enhanced via a hybrid-pooling graph attention network; functional features are processed similarly to textual features. (b) Early fusion stage: Modal features of drugs and proteins are aligned using Gram Loss, followed by integration through a multi-source cross-attention module. (c) Late fusion stage: A bidirectional cross-attention module captures fine-grained drug-target interactions, and a deep orthogonal fusion module reduces redundancy in fused features. (d) Classification head: The final prediction is made using an MLP classifier based on the fused drug-target representation.}
\label{fig:2} 
\end{figure*}

\textbf{Textual Feature Extraction:} textual features serve as the foundational modality describing drug molecules and protein sequences. We employ a multi-head self-attention (MHSA) module. This mechanism identifies semantic associations between key amino acids or critical semantic units in chemical descriptions, thereby enhancing the representation of long-range dependencies and contextual relations within the sequence. Consequently, it strengthens semantic expressiveness while maintaining stability and preserving raw information. We further enhance non-linear feature representations through a feed-forward network. 
Let the textual features be represented as $x_{t/f} \in \mathbb{R}^{Seq \times Dim}$, then the MHSA computation is given by eq~\eqref{eq:1}:
\begin{equation}
\begin{aligned}
    x_{t/f}^{head,i} &= \text{SoftMax}\left(\frac{W_h^Q x_{t/f} (W_h^K x_{t/f})^T}{\sqrt{d_k}}\right) \\
    &\times W_h^V x_{t/f}, \\
    x_{t/f}^{'} &= \left[x_{t/f}^{head,1} \| x_{t/f}^{head,2} \| \dots \| x_{t/f}^{head,n}\right], \\
    x_{t/f}^{'} &= \text{LN}(\text{MLP}(\text{LN}(x_{t/f}^{'}))), \\
\end{aligned}
\label{eq:1}
\end{equation}
Here, $\mathbb{R}^{Seq \times Dim}$ denotes a tensor with Seq sequence Dim dimensions, $x_{t/f}^{'} \in \mathbb{R}^{Seq \times Dim}$ denotes the textual/functional features after MHSA, while $Q$, $K$, $V \in \mathbb{R}^{Seq \times \frac{Dim}{H}}$ are the query, key, and value matrices. The parameters $W_h^Q$, $W_h^K$, $W_h^V \in \mathbb{R}^{Dim \times \frac{Dim}{H}}$ are learnable weights for the i-th attention head, $H$ is the total number of heads, and $d_k$ is the dimensionality of keys. The operator $\|$ indicates concatenation, $\text{Softmax}(\cdot)$ is the softmax function, $\text{LayerNorm}(\cdot)$ denotes layer normalization, and $\text{MLP}(\cdot)$ is the feed-forward network.

\textbf{Structural Feature Extraction:} In DTI prediction, structural features are critical for representing the graph topology of molecules and the spatial conformations of proteins. These features help the model capture the connectivity and interactions among atoms and amino acids. We design a mixed pooling graph attention module (MPGAT) to obtain holistic structural representations. This module applies multiple layers of graph attention networks (GATs) with residual connections to aggregate local neighborhood features and extract higher-order topological information. Afterward, diverse pooling strategies—including global max pooling, mean pooling, additive pooling, variance pooling, upper-quartile pooling, and lower-quartile pooling—are employed to form graph-level representations. Formally, let structural features be represented as $x_{g} \in \mathbb{R}^{(e,v)}$, the HPGAT computation can be expressed as eq~\eqref{eq:2}:
\begin{equation}
\begin{aligned}
x_{g}^{a} &= GAT(x_{g}) + x_{g}, \\
h_{max} &= \text{MaxPool}(x_{g}^{a}), \\
h_{avg} &= \text{AvgPool}(x_{g}^{a}), \\
h_{add} &= \text{AddPool}(x_{g}^{a}), \\
h_{var} &= \text{VarPool}(x_{g}^{a}), \\
h_{upq} &= \text{UpqPool}(x_{g}^{a}), \\
h_{lwq} &= \text{LwqPool}(x_{g}^{a}), \\
x_{g}^{'} &= \left[h_{max} \| h_{avg} \| h_{add} \| h_{var} \| h_{upq} \| h_{lwq}\right].
\end{aligned}
\label{eq:2}
\end{equation}
Here, $x_{g}^{'} \in \mathbb{R}^{N_{pool} \times Dim}$ denotes the structural feature after hybrid pooling, with $N_{pool}=6$. Each pooling function provides complementary perspectives of the graph structure, ensuring more comprehensive representation learning.

\textbf{Functional Feature Extraction:} Functional features provide high-level biological semantics regarding molecular functions, biological processes, and cellular components, thereby compensating for the limitations of sequence and structural features. The extraction method is consistent with that of textual features, the detail process is shown in appendies.
\subsubsection{Multi-level Feature Fusion}

\textbf{Early Fusion Strategy:} As shown in Fig~\ref{fig:2}(b), the early fusion strategy of the  CDI-DTI framework is illustrated. 
First, for the text, structure, and function modal features after feature extraction, we perform alignment operations. In conventional contrastive learning, traditional pairwise alignment methods based on cosine similarity always suffer from scalability bottlenecks. In contrast, we introduce the Gramian Representation Alignment Measure Loss (GRAM Loss) \cite{cicchetti2024gramian}, which measures the alignment degree between modalities by computing the volume of a k-dimensional parallelepiped. This method first performs unit normalization on the input modalities to ensure that the geometric computation is not affected by scale. The Gram matrix constructed subsequently contains the dot product information between modalities. By computing its determinant and taking the square root to obtain the geometric volume, and then minimizing it through cross-entropy loss, the model is indirectly driven to learn highly synergistic embedding representations across modalities. A smaller volume indicates that different modal features are closer to the same subspace, thereby reflecting a stronger alignment relationship. This study employs the volume-minimization-based Gram Loss mechanism to strengthen the mutual approximation of different modalities in the semantic space. The Gram Loss can be expressed as eq~\eqref{eq:3}:
\begin{equation}
\begin{aligned}
& x_{t/g/f}' = \frac{x_{t/g/f}'}{\sqrt{\sum_{j=1}^{Dim} x_{t/g/f,j}^2}},\\
& G = \begin{bmatrix}
x_{t}' x_{t}', & x_{t}' x_{g}', & x_{t}' x_{f}', \\
x_{g}' x_{t}', & x_{g}' x_{g}', & x_{g}' x_{f}', \\
x_{f}' x_{t}', & x_{f}' x_{g}', & x_{f}' x_{f}',\\
\end{bmatrix} \\
& V = \sqrt{\det(G) + \varepsilon}, \\
& \text{Gram Loss} = -\frac{1}{B} \sum_{i=1}^{B} \log \frac{\exp(-V / \tau)}{\sum_{j=1}^{k} \exp(-V_j / \tau)}.
\end{aligned} 
\label{eq:3}
\end{equation}
where $G \in \mathbb{R}^{Seq \times 3 \times 3}$ represents the feature similarity matrix, $\det(\cdot)$ denotes the volume computation function, $\varepsilon = 1 \times 10^{-8}$ is a small constant to prevent numerical instability due to a zero determinant, $V$ represents the volume of the parallelepiped formed by the features, and $\tau$ is the temperature parameter.
On the basis of modal feature alignment (minimizing volume) by Gram Loss, we utilize a Multi-source Cross-attention module to dynamically capture fine-grained interactions between text, structure, and function modalities, integrate complementary information, and enhance the semantic consistency and comprehensiveness of drug and protein features. Multi-source cross-attention allows one modality (text) to dynamically attend to relevant information from other modalities (structure, function), capturing fine-grained interactions between modalities. The multi-source cross-attention (MCA) can be expressed as eq~\eqref{eq:4}:
\begin{equation}
\begin{aligned}
& x_{t \And g} = x_{t}' || x_{g}', \\
& x_{g \And f} = x_{g}' || x_{f}', \\
& x_{f \And t} = x_{f}' || x_{t}', \\
& x_j^{head,i} = \text{Softmax}\left( \frac{{W_h^Q x_j} ({W_h^K x_k})^T}{\sqrt{d_k}} \right) {W_h^V x_k}, \\
& j = t, g, f \\
& k = g \And f, f \And t, t \And g, \\
& x_j' = [ x_j^{head,1} || x_j^{head,2} || \dots || x_j^{head,n} ], \\
& x = \text{Linear}(x_{t}' || x_{g}' || x_{f}'),
\end{aligned} 
\label{eq:4}
\end{equation}
where $x \in \mathbb{R}^{Seq \times Dim}$ is the feature after the multi-source cross-attention module, $Q_{j/k}^{head,i}, K_{j/k}^{head,i}, V_{j/k}^{head,i} \in \mathbb{R}^{Seq \times \frac{Dim}{H}}$ are the query, key, and value, $W_h^Q, W_h^K, W_h^V \in \mathbb{R}^{Dim \times \frac{Dim}{H}}$ are the learnable parameters for the $i$-th head, $H$ is the total number of heads, $d_k$ denotes the dimension of the key ($K_{j/k}^{head,i}$), $\text{Softmax}(\cdot)$ denotes the Softmax function, $||$ denotes the concatenation operation, and $\text{Linear}(\cdot)$ denotes the Linear function.
After multi-modal feature alignment and fusion, we obtain comprehensive and unified drug features and target features. Some studies use drug feature-guided collaborative attention to capture associations between drug features and protein features; however, improper learning may lead to overemphasis on drug features, thereby weakening protein features. This imbalance reduces the model's prediction capability. Our goal is to enhance the interaction capability between drug features and target features through bidirectional guidance to achieve better DTI tasks. To this end, we design a Bidirectional Cross-Attention(BCA) module to dynamically capture complex semantic relationships between drugs and proteins through bidirectional inter-modal interactions. This module provides comprehensive and unified embedding representations for the DTI prediction task by enabling bidirectional information flow between drug features and protein features. Let the drug and protein features be $d, t \in \mathbb{R}^{Seq \times Dim}$. The BCA can be expressed as eq~\eqref{eq:5}:
\begin{equation}
\begin{aligned}
& d^{head,i} = \text{Softmax}\left( \frac{{W_h^Q d} ({W_h^K d})^T}{\sqrt{d_k}} \right){W_h^V d},\\
& t^{head,i} = \text{Softmax}\left( \frac{{W_h^Q t} ({W_h^K t})^T}{\sqrt{d_k}} \right){W_h^V t},\\
& d' = [ d^{head,1} || d^{head,2} || \dots || d^{head,n} ],\\
& t' = [ t^{head,1} || t^{head,2} || \dots || t^{head,n} ],\\
& f_{early} = \text{Linear}(d' || t').
\end{aligned}
\label{eq:5}
\end{equation}
where $f_{early} \in \mathbb{R}^{Seq \times Dim}$ is the drug-target representation feature after fusion by the bidirectional cross-attention module, $Q_{d/t}^{head,i}, K_{d/t}^{head,i}, V_{d/t}^{head,i} \in \mathbb{R}^{Seq \times \frac{Dim}{H}}$ are the query, key, and value, $W_h^Q, W_h^K, W_h^V \in \mathbb{R}^{Dim \times \frac{Dim}{H}}$ are the learnable parameters for the $i$-th head, $H$ is the total number of heads, $d_k$ denotes the dimension of the key ($K_{d/t}^{head,i}$), $\text{Softmax}(\cdot)$ denotes the Softmax function, $||$ denotes the concatenation operation, and $\text{Linear}(\cdot)$ denotes the Linear function.
\begin{figure}
\centerline{\includegraphics[width=0.5\textwidth]{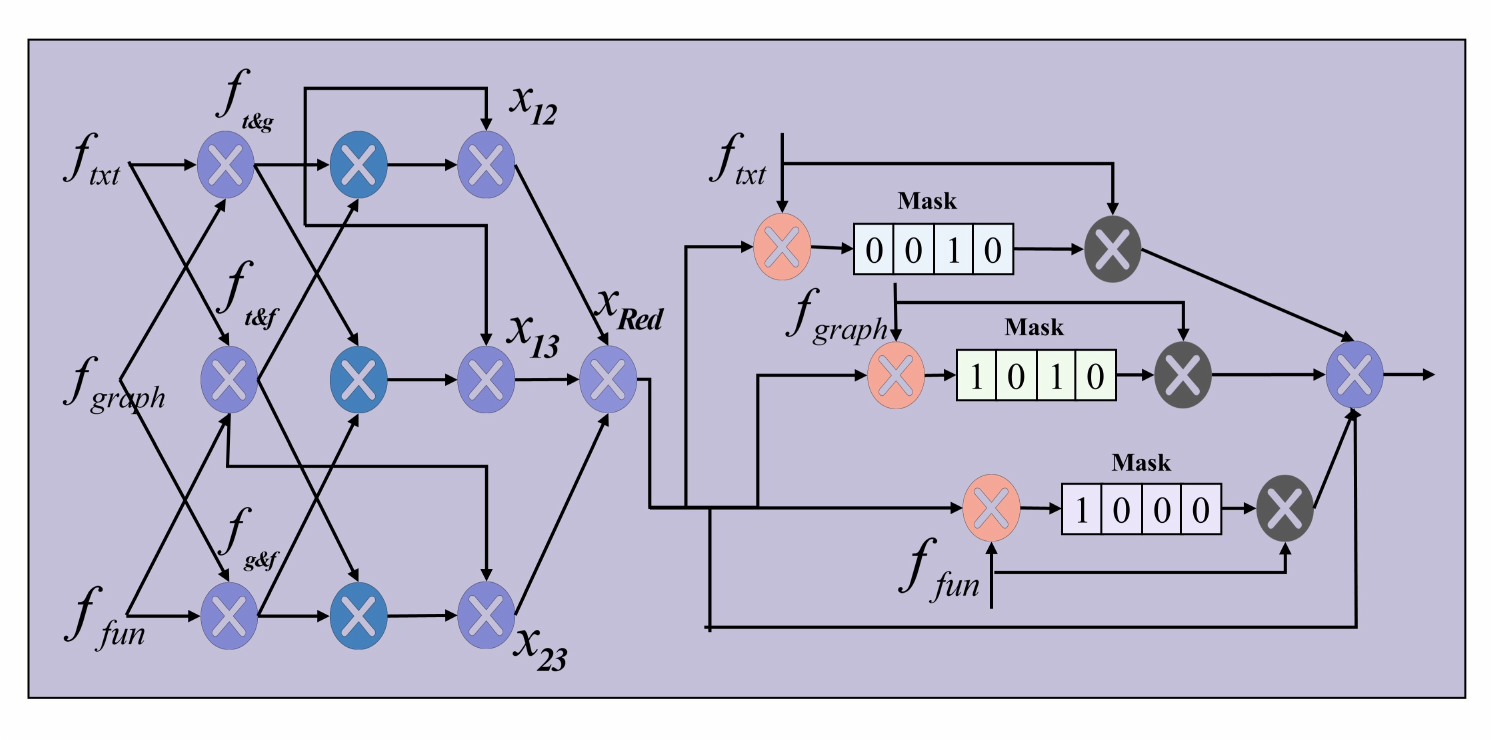}}
\caption{Deep Orthogonal Fusion (DOF) Module schematic diagram. This module constructs differential representations between modalities using linear transformations and orthogonal projections, followed by a masking mechanism to filter redundant features and enhance discriminative fused representations.}
\label{fig:3}
\end{figure}

\textbf{Late Fusion Stage:} Although the DTI features after early fusion can integrate comprehensive drug modal information and target modal information to form a unified interaction representation in the feature space, this fusion strategy may overlook fine-grained interaction information within modalities. Therefore, we design a late fusion stage to more effectively mine the fine interaction relationships between drugs and targets within modalities.
In the  CDI-DTI framework, the core idea of the late fusion strategy is: first, within modalities, through the BCA and Hybrid Graph Convolutional Fusion mechanism (HGCNF), deep interactions are performed on the three modal features of drugs and proteins to capture their potential associations from different perspectives, thereby generating three sets of high-quality DTI representation features. This process effectively enhances the synergistic expression capability between features while avoiding information bias issues that may arise from unidirectional attention.
Compared to the BCA module, we use the HGCNF module to fuse the structural modality by combining one-dimensional convolution and interaction matrix computation, which can effectively capture local patterns of features after pooling and global interaction information between drugs and proteins. The one-dimensional convolution operation aggregates local features through a sliding window, simulating local dependencies of neighborhood relationships in graphs, suitable for handling local chemical/biological patterns (such as bond connections or amino acid fragments) in drug molecules or protein sequences. At the same time, the interaction matrix models global interactions between drug and protein features through weighted modeling, enhancing cross-modal information fusion. This hybrid design retains the spatial structure of pooled features while considering global associativity, and can be expressed as eq~\eqref{eq:6}:
\begin{equation}
\begin{aligned}
& C = d_{g}' \cdot (t_{g}')^T, \quad C_T = t_{g}' \cdot (d_{g}')^T, \\
& f_d = \text{Linear}(d_{g}'), \quad f_t = \text{Linear}(t_{g}'), \\
& f_d' = \text{GNN}(f_d) + \text{Softmax}(C) f_d + f_d, \\
& f_t' = \text{GNN}(f_t) + \text{Softmax}(C_T) f_t + f_t, \\
& f_{g} = \text{Linear}(f_d' || f_t').
\end{aligned} \label{eq:6}
\end{equation}
where $\text{Linear}(\cdot)$ denotes the Linear function, $\text{GNN}(\cdot)$ denotes the one-dimensional convolution function, $\text{Softmax}(\cdot)$ denotes the Softmax function, and $||$ denotes the concatenation operation.
However, with the accumulation of dimensional stacking and redundant information from interactions of different modal features, traditional linear fusion struggles to sufficiently compress the expression space and may even introduce redundant interference.
To address this, we further introduce the Deep Orthogonal Fusion (DOF) module\cite{su2024mski} for fine integration of representations from different modal interactions in the late stage as shown in Figure~\ref{fig:3}. This module takes the three modal features as input, constructs differential representations between modalities through linear transformations and orthogonal projections, and on this basis, utilizes a masking mechanism for filtering and reconstruction of redundant features, further enhancing the discriminative ability and compactness of the fused features. It can be expressed as eq~\eqref{eq:7}:
\begin{equation}
\begin{aligned}
& f_{t \And g} = f_{t} || f_{g}, \\
& f_{t \And f} = f_{t} || f_{f}, \\
& f_{g \And f} = f_{g} || f_{f}, \\
& x_{12} = f_{t \And g} || (f_{t \And g} - \frac{f_{t \And g} f_{t \And f}}{||f_{t \And g}||^2} f_{t \And g}), \\
& x_{13} = f_{t \And g} || (f_{t \And g} - \frac{f_{t \And g} f_{g \And f}}{||f_{t \And g}||^2} f_{t \And g}), \\
& x_{23} = f_{t \And f} || (f_{t \And f} - \frac{f_{t \And f} f_{g \And f}}{||f_{t \And f}||^2} f_{t \And f}), \\
& x_{\text{Red}} = \text{Linear}(x_{12} || x_{13} || x_{23}), \\
& \text{mask(m)} = \mathbb{I} (\text{sim} > \text{threshold}).
\end{aligned} \label{eq:7}
\end{equation}
where $f_{later} \in \mathbb{R}^{Seq \times Dim}$ is the DTI feature representation after multi-modal late fusion, $x_{12}, x_{13}, x_{23}, x_{\text{Red}}$ are the redundant representations after orthogonal computation, $\mathbb{I}(\cdot)$ is the indicator function, and $\text{threshold}$ is a threshold used to filter out the most similar targets, avoiding redundancy and retaining specific information. The mask is computed as the similarity between the redundant representations and the DTI feature representations of different modalities, and a binary mask is generated based on the similarity scores. Similar to the attention mechanism, we utilize key-value pairs ($x_{\text{Red}}, f_m$) and query vector $f_m$ to compute cosine similarity:
\begin{equation}
\begin{aligned}
& \text{sim(m)} = \text{Cos}(tK(x_{\text{Red}}), tQ(f_m)), \\
& z_m = \text{mask(m)} \times tV(f_m), \quad m = t/g/f, \\
& f_{later} = \text{Linear}(z_{t} || z_{g} || z_{f} || x_{\text{Red}}).
\end{aligned} \label{eq:8}
\end{equation}
where $tQ(\cdot), tK(\cdot), tV(\cdot)$ denote linear transformations. Under the guidance of estimating similarity $\text{Sim}$, we filter out relatively similar features (high similarity scores) and only select features from specific modalities (low similarity scores) for final fusion.
\subsection{Interaction Prediction}
After the early fusion and late fusion stages, we obtain multi-modal DTI representations with different meanings at different stages: $f_{t}, f_{g}, f_{f}, f_{early}, f_{later}$. We integrate the early fusion features and late fusion features through linear concatenation:
\begin{align}
f_{output} = \text{Linear}(f_{early} || f_{later})
\label{eq:9}
\end{align}
where $f_{output} \in \mathbb{R}^{Seq \times Dim}$ is the integrated feature representation, and $\text{Linear}(\cdot)$ is the linear function. To optimize the network, we employ the cross-entropy loss function for DTI prediction,
The  CDI-DTI method proposed in this paper is a multi-task learning framework that includes 6 branches for extracting text, structure, and function features. We use different loss functions to train the 6 branch tasks, including $CE_{t}, CE_{g}, CE_{f}, CE_{early}, CE_{later}, CE_{output}$. Therefore, our total classification loss is defined as:
\begin{equation}
\begin{aligned}
   L_c &= \frac{1}{6} \left( CE_{t} + CE_{g} + CE_{f} + CE_{early} + CE_{later} \right) \\
       &\quad + CE_{output}
\end{aligned}
\label{eq:10}
\end{equation}
Our alignment loss Gram Loss includes the drug feature alignment loss $Gram_d$ and the target feature alignment loss $Gram_t$ in multi-modal, and the total alignment loss is defined as eq~\eqref{eq:11}:
\begin{align}
    L_g = (Gram_d + Gram_t) / 2 \label{eq:11}
\end{align}
Finally, we obtain the final loss function as eq~\eqref{eq:12}:
\begin{align}
    L = \frac{L_c + \lambda L_g}{1 + \lambda} 
\label{eq:12}
\end{align}
where $\lambda$ is the loss weight.
\section{RESULTS}
\subsection{Experimental Settings}
The initialization parameters of CDI-DTI are as follows: batch size of 32, Adam optimizer \cite{kingma2014adam} with a learning rate of 1e-4, learning rate decay interval of 10 epochs, weight decay coefficient of 1e-5, maximum training epochs of 50, hidden layer dimension of 512, number of graph self-attention layers of 3, number of multi-head self-attention layers of 1, loss weight of 1.0, and temperature parameter of 0.1. In this study, the proposed method uses the PyTorch library \cite{paszke2019pytorch} to conduct experiments, and the model is trained on NVIDIA RTX A100 40GB*1. We use accuracy, F1-score, AUROC (area under the receiver operating characteristic curve), and AUPRC (area under the precision-recall curve) as performance metrics. In addition, all specific tabular data for our experiments are provided in the appendix.
\subsection{Ablation Experiments}
\begin{figure*}[htbp]
\centering
\includegraphics[width=\textwidth]{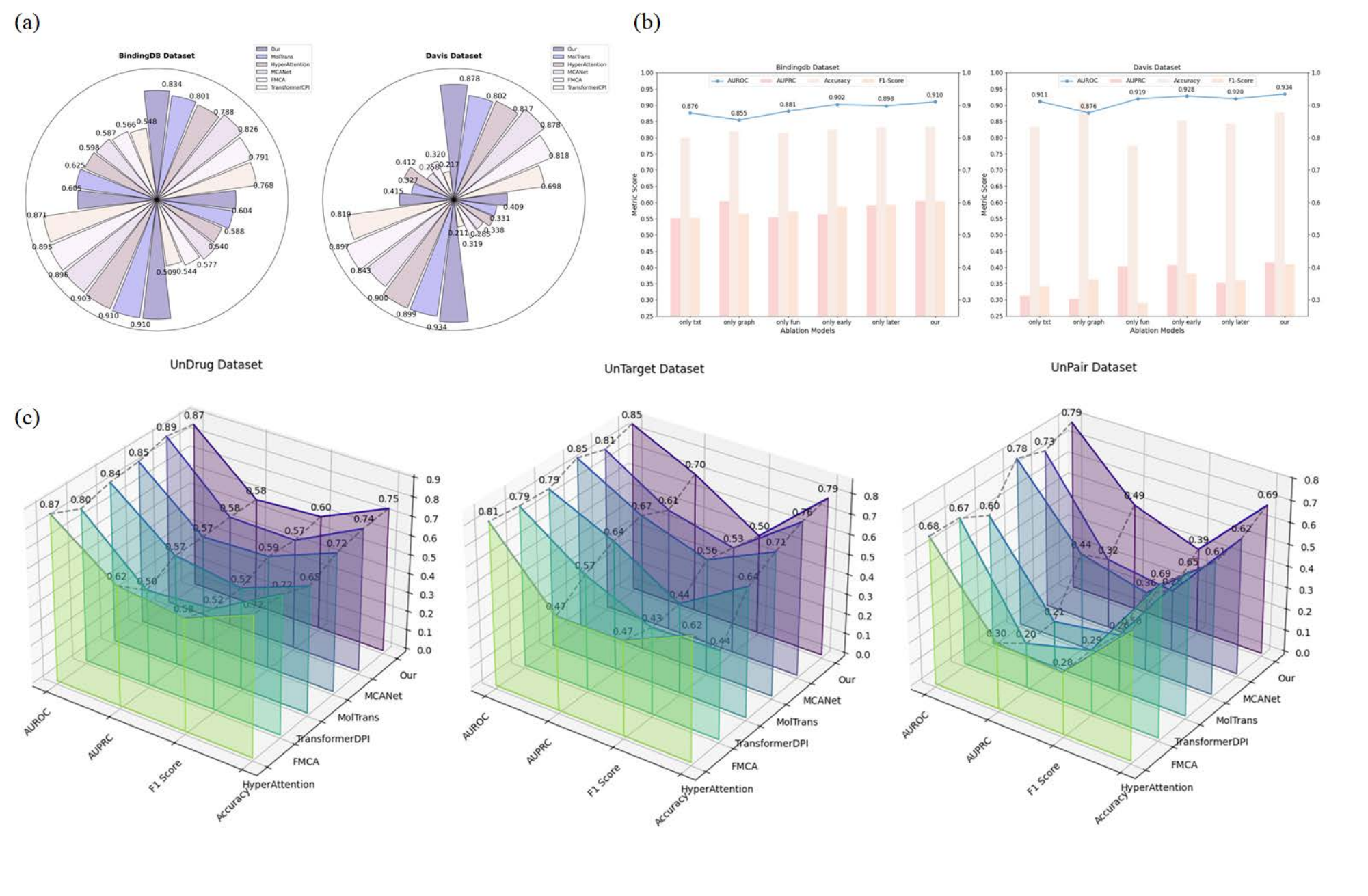}
\caption{The  CDI-DTI achieves accurate prediction of drug-target interactions. Below are the experimental results on BindingDB and Divas datasets. (a) Sector plots show the performance comparison of  CDI-DTI with other five state-of-the-art methods (b) bar plots show the experimental results of ablation for each modality of  CDI-DTI (c) 3D line plots show the performance comparison of  CDI-DTI with other five state-of-the-art methods in the cold start experiment.}
\label{fig:results}
\end{figure*}

To comprehensively evaluate the contribution of each module in the proposed  CDI-DTI framework, we conducted ablation experiments on the BindingDB and Davis datasets. We verified the roles of the four core modules: multi-modal feature extraction, early fusion, late fusion, and interaction prediction head, including three single-modal inputs (Only txt, Only graph, Only fun), two fusion strategies (Only early, Only later), and the complete  CDI-DTI model (our). The ablation experiment results are shown in Figure~\ref{fig:results}(b).

The results indicate that all single-modality settings perform significantly worse than the full model, confirming the complementarity among modalities and the inadequacy of relying on a single modality to fully characterize the complex interactions between drugs and targets. Further analysis of fusion strategies reveals that the variant using only early fusion (Only early) significantly outperforms any single-modality model on both datasets. On BindingDB, AUROC improves to 0.9025, and Accuracy reaches 0.8242; however, this strategy still falls short of the full model(AUROC: 0.9104), indicating that early semantic alignment alone cannot fully capture fine-grained interaction relationships. Conversely, the variant skipping early fusion and retaining only late fusion (Only later) uses BCA and DOF modules to integrat DTI features from three modalities at the decision level, still outperforming single-modality models but underperforming the full model due to the lack of early inter-modal alignment. Overall, the complete  CDI-DTI model integrates Multi-modal feature extraction and mutil-stage fusion, achieving optimal performance across all metrics on both datasets. This validates the effectiveness of the collaborative design of its components, particularly the early and late fusion strategies, which play a critical role in capturing complementary relationships at different levels in Multi-modal features. These results confirm the rationality and superiority of the proposed Multi-modal multi-stage fusion mechanism in DTI prediction tasks.
\subsection{Single-Domain Experiments}
To comprehensively evaluate the effectiveness of  CDI-DTI, we conducted comparative experiments with mainstream methods on two public datasets, BindingDB and Davis. The results, shown in Fig.~\ref{fig:results}(a), demonstrate that our method outperforms others in multiple key metrics—Accuracy, F1-score, AUROC, and AUPRC—validating its significant advantages in Multi-modal modeling and interaction fusion. Specifically, on the BindingDB dataset, our method achieves superior Accuracy (0.8336), F1-score (0.604), and AUROC (0.9104) compared to strong baselines like MolTrans\cite{huang2021moltrans}, MCANet\cite{bian2023mcanet}, HyperAttention\cite{zhao2022hyperattentiondti}, TransformerCPI\cite{chen2020transformercpi} and FMCA\cite{zhang2024fmca}, with particularly stable F1-score performance, indicating robust generalization under imbalanced sample distributions. In AUPRC, our method ranks just behind MolTrans but maintains the most balanced overall metrics, demonstrating the ability to maintain high recall while accurately identifying positive classes. On the Davis dataset, our method continues to lead, achieving the best scores in Accuracy (0.8777), F1-score (0.4085), AUROC (0.9340), and AUPRC (0.4149), with notable improvements in F1-score and AUPRC, highlighting its modeling capability in small-sample scenarios. Our method not only exhibits strong performance stability across datasets with different distributions but also confirms its broad adaptability and practical potential in Multi-modal DTI prediction tasks.
\subsection{Cross-Domain Experiments}
To further validate the model's generalization across different data domains, we conducted cross-domain transfer experiments between the BindingDB and Davis datasets, using two transfer strategies: B→D (trained on BindingDB, tested on Davis) and D→B (trained on Davis, tested on BindingDB). The results, as shown in Table~\ref{tab:3}, indicate that our method significantly outperforms mainstream models in both transfer directions, demonstrating stronger cross-domain robustness across four key metrics: Accuracy, F1-score, AUROC, and AUPRC.

In the B→D experiment, our method achieves the best Accuracy (0.7706) and AUROC (0.8118), with a clear improvement in F1-score (0.2292) compared to other methods. Although the AUPRC is slightly lower than TransformerDPI, our method's overall performance is the most balanced. Notably, traditional methods like MolTrans and HyperAttention exhibit significant performance degradation in this transfer scenario, indicating sensitivity to domain distribution changes, whereas our method maintains stable discriminative ability, proving stronger transferability. In the D→B experiment, our method again leads in Accuracy (0.7989) and F1-score (0.3223), with robust performance in AUROC and AUPRC, significantly outperforming other methods. Most methods experience substantial performance degradation in this task, particularly in AUPRC, while our model remains competitive. This cross-domain robustness stems from our method's structural advantages, leveraging three heterogeneous modality features to comprehensively characterize drugs and protein sequences. Our method achieves stable and superior performance in both cross-domain transfer tasks, further validating its effectiveness and practical value in DTI prediction across domains.
\begin{table*}[htbp]
\centering
\caption{Cross-Domain Comparison Experiment Results}
\begin{tabular*}{\textwidth}{l@{\extracolsep{\fill}}|cccc|cccc}
\toprule
Model/Dataset & \multicolumn{4}{c}{B$\to$D} & \multicolumn{4}{c}{D$\to$B} \\
 & Accurary & F1-score & AUROC & AUPRC & Accurary & F1-score & AUROC & AUPRC \\
\midrule
FMCA & 0.7705 & 0.1758 & 0.6951 & 0.1383 & 0.7454 & 0.3132 & 0.6975 & 0.2691 \\
TransformerCPI & 0.7058 & 0.1865 & 0.7366 & 0.1842 & 0.6237 & 0.3160 & 0.6926 & 0.3125 \\
MolTrans & 0.4303 & 0.1521 & 0.7112 & 0.1236 & 0.6473 & 0.1580 & 0.6473 & 0.2348 \\
MACNet & 0.7090 & 0.1865 & 0.6862 & 0.1227 & 0.8054 & 0.2221 & 0.5990 & 0.2686 \\
HyperAttention & 0.4373 & 0.1189 & 0.6268 & 0.1010 & 0.7725 & 0.1788 & 0.5864 & 0.1805 \\
Our & 0.7706 & 0.2292 & 0.8118 & 0.1858 & 0.7989 & 0.3223 & 0.6984 & 0.2746 \\
\bottomrule
\end{tabular*}
\label{tab:3}
\end{table*}
\subsection{Cold-Start Scenario Experiments}
To evaluate the generalization ability of our model in cold-start scenarios, we designed three challenging tasks: unDrug (unseen new drugs), unTarget (unseen new targets), and unPair (novel drug-target pairs). These settings are common in real-world drug discovery, where the compounds or targets to be predicted are absent from the training set, posing higher demands on the model's generalization and inference capabilities. The experimental results, as shown in Fig~\ref{fig:results}(c), demonstrate that our method exhibits superior performance across all three subtasks, achieving higher consistency and robustness in comprehensive metrics (Accuracy, F1, AUROC, and AUPRC).

In the unDrug task, our method outperforms existing approaches in Accuracy (0.7502) and F1-score (0.5972), while also leading in AUROC (0.8672). This indicates that our method effectively captures the semantic and structural features of new drug molecules, enabling high-quality predictions. Although the AUPRC is slightly lower than some methods, the overall evaluation highlights our method's stability and expressiveness in drug-level cold-start scenarios. The unTarget task further validates our model's generalization at the target level, with significant leads in Accuracy (0.7875) and AUPRC (0.6992), and the best AUROC (0.8515) for this task. This performance underscores the effectiveness of our method in characterizing protein modalities, largely due to the introduction of functional modalities, which, for the first time, incorporate protein biological function attributes into DTI modeling, enhancing adaptability to unseen targets. In the most challenging unPair task, where the model faces entirely unseen drug-target pairs, our method maintains stable performance, with leading AUROC (0.7854) and AUPRC (0.4942), and a notably improved F1-score (0.3860) compared to existing methods. The superior performance in cold-start tasks stems from the structural design of our model, which exhibits structural robustness and semantic generalization, resulting in significant performance improvements across all three cold-start tasks.
\subsection{Visualization Experiments}
\begin{figure}
\centering
\includegraphics[width=0.5\textwidth]{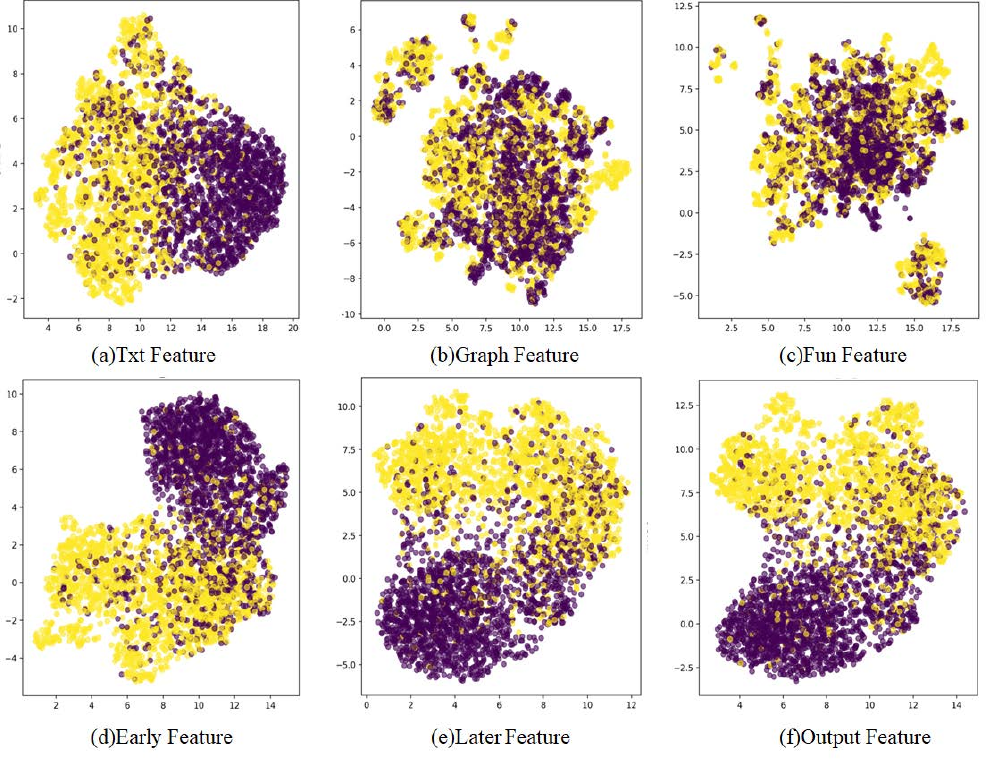}
\caption{Visualization of Different Modal Features and Fusion Representations at Each Stage. (a) Text Modality Features. (b) Graph Structural Modality Features. (c) Functional Modality Features. (d) Early Fusion Features. (e) Late Fusion Features. (f) Final Output Features.}
\label{fig:visualization}   
\end{figure}

In order to deeply analyze the performance of the model in the process of multi-modal feature extraction and fusion, UMAP dimensionality reduction visualization is performed on different modal features and fusion representations at each stage. The result is shown in Fig.~\ref{fig:visualization}. The yellow and purple points in each subfigure represent positive and negative samples, respectively, and are used to observe the separability of the samples in the embedding space. From the perspective of single-modal feature distribution, the necessity of multi-modal fusion is preliminarily verified. In the early fusion stage (d), the modality information shows more obvious intra-class aggregation and inter-class separation, indicating that the early fusion mechanism effectively aligns the semantic differences between heterogeneous modalities. In the late fusion representation (e), the category boundaries of the samples are clearer and the concentration of negative samples (purple) is improved. The final output feature (f) shows an optimal clustering structure, positive and negative samples form a clear distribution boundary in the embedding space, and the overlap area between classes is significantly reduced. This phenomenon confirms the effect of layer-by-layer optimization of the whole modeling process on the representation layer, and also confirms the effectiveness of the proposed multi-stage fusion mechanism in capturing complex molecular-protein interaction patterns.
\section{Discussion}
\begin{figure*}
\centering
\includegraphics[width=\textwidth]{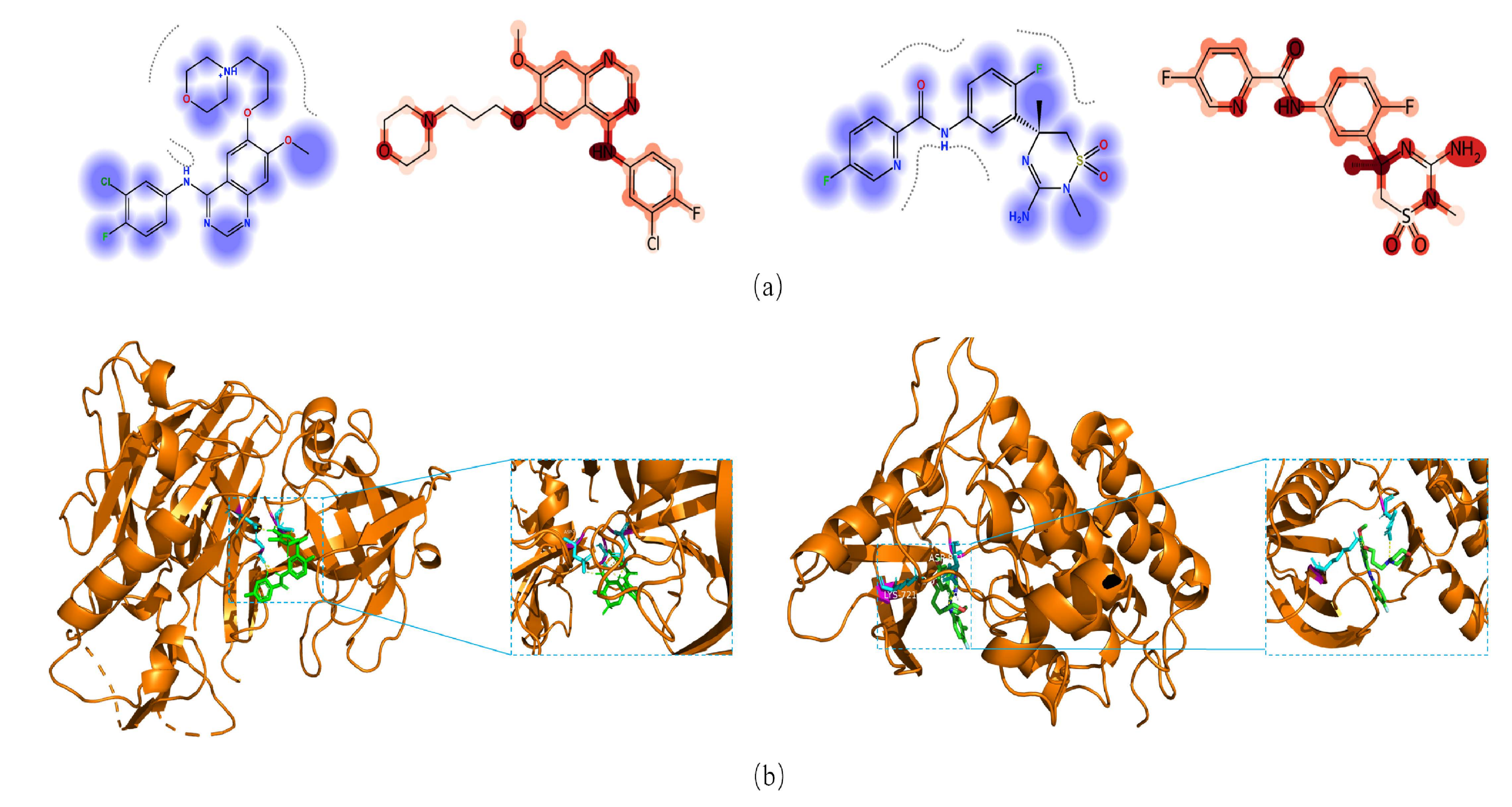}
\caption{Interpretable Learning for CDI-DTI. a, Interpretability of co-crystallized ligands. The 2D structure of the ligand is shown on the left side of each figure. The right side of each panel shows the importance of each atom. b, Interpretability of ligand-receptor docking residues. All ligand-protein interaction maps and 3D structures can be visualized by AutoDockTool.}    
\label{fig:interpretability}
\end{figure*}
In DTI prediction research, interpretability has always been a key issue for whether artificial intelligence methods can truly land in drug discovery practices. Traditional deep learning models, although achieving high prediction accuracy, are often regarded as "black boxes," making it difficult to reveal their internal decision mechanisms. The CDI-DTI model, through a multi-granularity multi-attention fusion mechanism, not only improves prediction performance but also provides an intuitive way to parse the interaction mechanisms between ligands and targets. Specifically, the atomic-level attention weights generated by the model can annotate key groups on the two-dimensional molecular structure as shown in Fig.~\ref{fig:interpretability}(a) and form mutual verification with three-dimensional binding patterns obtained from molecular docking and crystal structure analysis as shown in Fig.~\ref{fig:interpretability}(b). This combination of "interpretable machine learning" and "structural biology evidence" makes DTI modeling go beyond numerical prediction and provide molecular interaction maps closer to pharmacological empirical evidence.

We use two classic drug-target systems as examples to verify the rationality of the CDI-DTI model's attention weights. First, for epidermal growth factor receptor (EGFR, UniProtID: P00533) and its inhibitor Gefitinib (DrugBankID: DB00317), the model's attention heatmap shows significant distribution on the quinazoline ring N1, aromatic ring, and adjacent substituents of Gefitinib, completely consistent with MOE molecular docking results. Specifically, the key residues identified by the model include Met793 forming a hydrogen bond with quinazoline ring N1 , Leu718, and Val726 providing hydrophobic pocket interactions , which highly match the reported Gefitinib binding mode in the literature \cite{lynch2004activating}. This result indicates that CDI-DTI can not only capture the overall trends of DTI but also localize specific atomic-residue interaction interfaces. Second, for $\beta$-secretase 1 (BACE1, UniProtID: P56817) and its inhibitor Verubecestat (DrugBankID: DB12130), the model's attention map highlights its polar amino groups and nitrogen-containing aromatic ring regions. AutoDock docking results show that Verubecestat forms hydrogen bonds with Asp32 and Asp228 in the BACE1 active pocket and interacts with Gly230 to stabilize the region, consistent with the experimental resolution of the BACE1 crystal structure\cite{novak2020long}. The attention distribution of CDI-DTI highly overlaps with these docking and literature evidence, further verifying the model's reliability in revealing real molecular interaction mechanisms.

The interpretability of CDI-DTI lies not only in verifying known interaction mechanisms but more importantly in its guiding value for future drug design. Taking EGFR as an example, Met793 is a key binding residue for a class of clinical EGFR inhibitors such as Osimertinib , and its mutation like T790M often leads to resistance \cite{yun2008t790m}. The model's ability to automatically highlight this site indicates its potential for resistance mechanism research and the design of next-generation inhibitors. Similarly, the Asp32/Asp228 dual acidic groups of BACE1 are known pharmacophore core regions, with almost all BACE1 inhibitors relying on hydrogen bonding with them. The model's correct capture of this pattern indicates its ability to identify core pharmacological action units. These results not only enhance the interpretability of predictions but also provide specific structural targets for molecular modification and new drug optimization.

\section{Conclusion}
In this work, we propose CDI-DTI, captures the complementary nature of different data modalities, significantly enhancing prediction accuracy and generalization. Extensive experiments on BindingDB and Davis datasets demonstrate that CDI-DTI consistently outperforms or achieves near-optimal performance compared to mainstream baselines. Additionally, cross-domain and cold-start evaluations reveal that CDI-DTI exhibits minimal performance degradation when faced with unseen drugs, targets, or pairs, highlighting its robustness and ability to maintain stability across various domains.

Moreover, CDI-DTI offers valuable interpretability through its attention mechanism, providing insights into the molecular interaction process. For example, in the EGFR-Gefitinib and BACE1-Verubecestat systems, the model identifies key residues such as Met793 of EGFR and Asp32/Asp228 of BACE1, which align with experimental findings, enhancing our understanding of the underlying interaction mechanisms. This interpretability makes CDI-DTI particularly suitable for real-world applications in drug optimization, resistance analysis, and the repurposing of existing drugs.

Looking forward, future work will focus on scaling to larger datasets and further experimental validation. In summary, CDI-DTI advances beyond traditional black-box DTI models by balancing predictive performance, cross-domain generalization, and interpretability. As explainable AI continues to evolve, it offers the potential for a more efficient, transparent pipeline in precision medicine and the development of novel therapies.
\section{Acknowledgment}
The authors declare no competing interests.
\section*{References}
\bibliographystyle{IEEEtran}
\bibliography{myref}

\begin{thebibliography}{10}
\providecommand{\url}[1]{#1}
\csname url@samestyle\endcsname
\providecommand{\newblock}{\relax}
\providecommand{\bibinfo}[2]{#2}
\providecommand{\BIBentrySTDinterwordspacing}{\spaceskip=0pt\relax}
\providecommand{\BIBentryALTinterwordstretchfactor}{4}
\providecommand{\BIBentryALTinterwordspacing}{\spaceskip=\fontdimen2\font plus
\BIBentryALTinterwordstretchfactor\fontdimen3\font minus \fontdimen4\font\relax}
\providecommand{\BIBforeignlanguage}[2]{{%
\expandafter\ifx\csname l@#1\endcsname\relax
\typeout{** WARNING: IEEEtran.bst: No hyphenation pattern has been}%
\typeout{** loaded for the language `#1'. Using the pattern for}%
\typeout{** the default language instead.}%
\else
\language=\csname l@#1\endcsname
\fi
#2}}
\providecommand{\BIBdecl}{\relax}
\BIBdecl

\bibitem{chen2016drug}
X.~Chen, C.~C. Yan, X.~Zhang, X.~Zhang, F.~Dai, J.~Yin, and Y.~Zhang, ``Drug--target interaction prediction: databases, web servers and computational models,'' \emph{Briefings in bioinformatics}, vol.~17, no.~4, pp. 696--712, 2016.

\bibitem{wen2017deep}
M.~Wen, Z.~Zhang, S.~Niu, H.~Sha, R.~Yang, Y.~Yun, and H.~Lu, ``Deep-learning-based drug--target interaction prediction,'' \emph{Journal of proteome research}, vol.~16, no.~4, pp. 1401--1409, 2017.

\bibitem{chen2018machine}
R.~Chen, X.~Liu, S.~Jin, J.~Lin, and J.~Liu, ``Machine learning for drug-target interaction prediction,'' \emph{Molecules}, vol.~23, no.~9, p. 2208, 2018.

\bibitem{dimasi2003price}
J.~A. DiMasi, R.~W. Hansen, and H.~G. Grabowski, ``The price of innovation: new estimates of drug development costs,'' \emph{Journal of health economics}, vol.~22, no.~2, pp. 151--185, 2003.

\bibitem{paul2010improve}
S.~M. Paul, D.~S. Mytelka, C.~T. Dunwiddie, C.~C. Persinger, B.~H. Munos, S.~R. Lindborg, and A.~L. Schacht, ``How to improve r\&d productivity: the pharmaceutical industry's grand challenge,'' \emph{Nature reviews Drug discovery}, vol.~9, no.~3, pp. 203--214, 2010.

\bibitem{pushpakom2019drug}
S.~Pushpakom, F.~Iorio, P.~A. Eyers, K.~J. Escott, S.~Hopper, A.~Wells, A.~Doig, T.~Guilliams, J.~Latimer, C.~McNamee \emph{et~al.}, ``Drug repurposing: progress, challenges and recommendations,'' \emph{Nature reviews Drug discovery}, vol.~18, no.~1, pp. 41--58, 2019.

\bibitem{zhang2022deepmgt}
P.~Zhang, Z.~Wei, C.~Che, and B.~Jin, ``Deepmgt-dti: Transformer network incorporating multilayer graph information for drug--target interaction prediction,'' \emph{Computers in biology and medicine}, vol. 142, p. 105214, 2022.

\bibitem{bagherian2021machine}
M.~Bagherian, E.~Sabeti, K.~Wang, M.~A. Sartor, Z.~Nikolovska-Coleska, and K.~Najarian, ``Machine learning approaches and databases for prediction of drug--target interaction: a survey paper,'' \emph{Briefings in bioinformatics}, vol.~22, no.~1, pp. 247--269, 2021.

\bibitem{ding2014similarity}
H.~Ding, I.~Takigawa, H.~Mamitsuka, and S.~Zhu, ``Similarity-based machine learning methods for predicting drug--target interactions: a brief review,'' \emph{Briefings in bioinformatics}, vol.~15, no.~5, pp. 734--747, 2014.

\bibitem{wang2023fusion}
K.~Wang and M.~Li, ``Fusion-based deep learning architecture for detecting drug-target binding affinity using target and drug sequence and structure,'' \emph{IEEE Journal of Biomedical and Health Informatics}, vol.~27, no.~12, pp. 6112--6120, 2023.

\bibitem{nguyen2021graphdta}
T.~Nguyen, H.~Le, T.~P. Quinn, T.~Nguyen, T.~D. Le, and S.~Venkatesh, ``Graphdta: predicting drug--target binding affinity with graph neural networks,'' \emph{Bioinformatics}, vol.~37, no.~8, pp. 1140--1147, 2021.

\bibitem{ozturk2018deepdta}
H.~{\"O}zt{\"u}rk, A.~{\"O}zg{\"u}r, and E.~Ozkirimli, ``Deepdta: deep drug--target binding affinity prediction,'' \emph{Bioinformatics}, vol.~34, no.~17, pp. i821--i829, 2018.

\bibitem{lee2019deepconv}
I.~Lee, J.~Keum, and H.~Nam, ``Deepconv-dti: Prediction of drug-target interactions via deep learning with convolution on protein sequences,'' \emph{PLoS computational biology}, vol.~15, no.~6, p. e1007129, 2019.

\bibitem{tsubaki2019compound}
M.~Tsubaki, K.~Tomii, and J.~Sese, ``Compound--protein interaction prediction with end-to-end learning of neural networks for graphs and sequences,'' \emph{Bioinformatics}, vol.~35, no.~2, pp. 309--318, 2019.

\bibitem{huang2021moltrans}
K.~Huang, C.~Xiao, L.~M. Glass, and J.~Sun, ``Moltrans: molecular interaction transformer for drug--target interaction prediction,'' \emph{Bioinformatics}, vol.~37, no.~6, pp. 830--836, 2021.

\bibitem{zhao2022hyperattentiondti}
Q.~Zhao, H.~Zhao, K.~Zheng, and J.~Wang, ``Hyperattentiondti: improving drug--protein interaction prediction by sequence-based deep learning with attention mechanism,'' \emph{Bioinformatics}, vol.~38, no.~3, pp. 655--662, 2022.

\bibitem{zeng2021deep}
Y.~Zeng, X.~Chen, Y.~Luo, X.~Li, and D.~Peng, ``Deep drug-target binding affinity prediction with multiple attention blocks,'' \emph{Briefings in bioinformatics}, vol.~22, no.~5, p. bbab117, 2021.

\bibitem{bian2023mcanet}
J.~Bian, X.~Zhang, X.~Zhang, D.~Xu, and G.~Wang, ``Mcanet: shared-weight-based multiheadcrossattention network for drug--target interaction prediction,'' \emph{Briefings in Bioinformatics}, vol.~24, no.~2, p. bbad082, 2023.

\bibitem{cheng2022iifdti}
Z.~Cheng, Q.~Zhao, Y.~Li, and J.~Wang, ``Iifdti: predicting drug--target interactions through interactive and independent features based on attention mechanism,'' \emph{Bioinformatics}, vol.~38, no.~17, pp. 4153--4161, 2022.

\bibitem{lee2024dlm}
J.~Lee, D.~W. Jun, I.~Song, and Y.~Kim, ``Dlm-dti: a dual language model for the prediction of drug-target interaction with hint-based learning,'' \emph{Journal of Cheminformatics}, vol.~16, no.~1, p.~14, 2024.

\bibitem{chen2020transformercpi}
L.~Chen, X.~Tan, D.~Wang, F.~Zhong, X.~Liu, T.~Yang, X.~Luo, K.~Chen, H.~Jiang, and M.~Zheng, ``Transformercpi: improving compound--protein interaction prediction by sequence-based deep learning with self-attention mechanism and label reversal experiments,'' \emph{Bioinformatics}, vol.~36, no.~16, pp. 4406--4414, 2020.

\bibitem{lin2020deepgs}
X.~Lin, K.~Zhao, T.~Xiao, Z.~Quan, Z.-J. Wang, and P.~S. Yu, ``Deepgs: Deep representation learning of graphs and sequences for drug-target binding affinity prediction,'' in \emph{ECAI 2020}.\hskip 1em plus 0.5em minus 0.4em\relax IOS Press, 2020, pp. 1301--1308.

\bibitem{ye2022molecular}
X.-b. Ye, Q.~Guan, W.~Luo, L.~Fang, Z.-R. Lai, and J.~Wang, ``Molecular substructure graph attention network for molecular property identification in drug discovery,'' \emph{Pattern Recognition}, vol. 128, p. 108659, 2022.

\bibitem{zhou2021multidti}
D.~Zhou, Z.~Xu, W.~Li, X.~Xie, and S.~Peng, ``Multidti: drug--target interaction prediction based on multi-modal representation learning to bridge the gap between new chemical entities and known heterogeneous network,'' \emph{Bioinformatics}, vol.~37, no.~23, pp. 4485--4492, 2021.

\bibitem{li2024mfcm}
W.~Li, W.~Ma, M.~Yang, and X.~Tang, ``Mfcm-dti model of multimodal feature fusion: prediction of drug-target interaction,'' in \emph{2024 IEEE International Conference on Bioinformatics and Biomedicine (BIBM)}.\hskip 1em plus 0.5em minus 0.4em\relax IEEE, 2024, pp. 687--692.

\bibitem{hu2025multi}
J.~Hu, Y.~Liu, X.~Zeng, Q.~Zou, R.~Su, and L.~Wei, ``Multi-modal deep representation learning accurately identifies and interprets drug-target interactions,'' \emph{IEEE Journal of Biomedical and Health Informatics}, 2025.

\bibitem{hua2025mmdg}
Y.~Hua, Z.~Feng, X.~Song, X.-J. Wu, and J.~Kittler, ``Mmdg-dti: Drug--target interaction prediction via multimodal feature fusion and domain generalization,'' \emph{Pattern Recognition}, vol. 157, p. 110887, 2025.

\bibitem{du2022compound}
B.-X. Du, Y.~Qin, Y.-F. Jiang, Y.~Xu, S.-M. Yiu, H.~Yu, and J.-Y. Shi, ``Compound--protein interaction prediction by deep learning: databases, descriptors and models,'' \emph{Drug discovery today}, vol.~27, no.~5, pp. 1350--1366, 2022.

\bibitem{li2025m3st}
X.~Li, R.~Su, and L.~Liu, ``M3st-dti: A multi-task learning model for drug-target interactions based on multi-modal features and multi-stage alignment,'' \emph{arXiv preprint arXiv:2510.12445}, 2025.

\bibitem{liu2007bindingdb}
T.~Liu, Y.~Lin, X.~Wen, R.~N. Jorissen, and M.~K. Gilson, ``Bindingdb: a web-accessible database of experimentally determined protein--ligand binding affinities,'' \emph{Nucleic acids research}, vol.~35, no. suppl\_1, pp. D198--D201, 2007.

\bibitem{davis2011comprehensive}
M.~I. Davis, J.~P. Hunt, S.~Herrgard, P.~Ciceri, L.~M. Wodicka, G.~Pallares, M.~Hocker, D.~K. Treiber, and P.~P. Zarrinkar, ``Comprehensive analysis of kinase inhibitor selectivity,'' \emph{Nature biotechnology}, vol.~29, no.~11, pp. 1046--1051, 2011.

\bibitem{chithrananda2020chemberta}
S.~Chithrananda, G.~Grand, and B.~Ramsundar, ``Chemberta: large-scale self-supervised pretraining for molecular property prediction,'' \emph{arXiv preprint arXiv:2010.09885}, 2020.

\bibitem{lin2022language}
Z.~Lin, H.~Akin, R.~Rao, B.~Hie, Z.~Zhu, W.~Lu, A.~dos Santos~Costa, M.~Fazel-Zarandi, T.~Sercu, S.~Candido \emph{et~al.}, ``Language models of protein sequences at the scale of evolution enable accurate structure prediction,'' \emph{BioRxiv}, vol. 2022, p. 500902, 2022.

\bibitem{jumper2021highly}
J.~Jumper, R.~Evans, A.~Pritzel, T.~Green, M.~Figurnov, O.~Ronneberger, K.~Tunyasuvunakool, R.~Bates, A.~{\v{Z}}{\'\i}dek, A.~Potapenko \emph{et~al.}, ``Highly accurate protein structure prediction with alphafold,'' \emph{nature}, vol. 596, no. 7873, pp. 583--589, 2021.

\bibitem{kulmanov2018deepgo}
M.~Kulmanov, M.~A. Khan, and R.~Hoehndorf, ``Deepgo: predicting protein functions from sequence and interactions using a deep ontology-aware classifier,'' \emph{Bioinformatics}, vol.~34, no.~4, pp. 660--668, 2018.

\bibitem{lee2020biobert}
J.~Lee, W.~Yoon, S.~Kim, D.~Kim, S.~Kim, C.~H. So, and J.~Kang, ``Biobert: a pre-trained biomedical language representation model for biomedical text mining,'' \emph{Bioinformatics}, vol.~36, no.~4, pp. 1234--1240, 2020.

\bibitem{edwards2022translation}
C.~Edwards, T.~Lai, K.~Ros, G.~Honke, K.~Cho, and H.~Ji, ``Translation between molecules and natural language,'' \emph{arXiv preprint arXiv:2204.11817}, 2022.

\bibitem{kang2022fine}
H.~Kang, S.~Goo, H.~Lee, J.-w. Chae, H.-y. Yun, and S.~Jung, ``Fine-tuning of bert model to accurately predict drug--target interactions,'' \emph{Pharmaceutics}, vol.~14, no.~8, p. 1710, 2022.

\bibitem{cicchetti2024gramian}
G.~Cicchetti, E.~Grassucci, L.~Sigillo, and D.~Comminiello, ``Gramian multimodal representation learning and alignment,'' \emph{arXiv preprint arXiv:2412.11959}, 2024.

\bibitem{su2024mski}
R.~Su, J.~Xiao, H.~Cui, P.~Xuan, X.~Feng, L.~Wei, and Q.~Jin, ``Mski-net: Towards modality-specific knowledge interaction for glioma survival prediction,'' in \emph{2024 IEEE International Conference on Bioinformatics and Biomedicine (BIBM)}.\hskip 1em plus 0.5em minus 0.4em\relax IEEE, 2024, pp. 2438--2445.

\bibitem{kingma2014adam}
D.~P. Kingma and J.~Ba, ``Adam: A method for stochastic optimization,'' \emph{arXiv preprint arXiv:1412.6980}, 2014.

\bibitem{paszke2019pytorch}
A.~Paszke, ``Pytorch: An imperative style, high-performance deep learning library,'' \emph{arXiv preprint arXiv:1912.01703}, 2019.

\bibitem{zhang2024fmca}
Q.~Zhang, L.~Zuo, Y.~Ren, S.~Wang, W.~Wang, L.~Ma, J.~Zhang, and B.~Xia, ``Fmca-dti: a fragment-oriented method based on a multihead cross attention mechanism to improve drug--target interaction prediction,'' \emph{Bioinformatics}, vol.~40, no.~6, p. btae347, 2024.

\bibitem{lynch2004activating}
T.~J. Lynch, D.~W. Bell, R.~Sordella, S.~Gurubhagavatula, R.~A. Okimoto, B.~W. Brannigan, P.~L. Harris, S.~M. Haserlat, J.~G. Supko, F.~G. Haluska \emph{et~al.}, ``Activating mutations in the epidermal growth factor receptor underlying responsiveness of non--small-cell lung cancer to gefitinib,'' \emph{New England Journal of Medicine}, vol. 350, no.~21, pp. 2129--2139, 2004.

\bibitem{novak2020long}
G.~Novak, J.~R. Streffer, M.~Timmers, D.~Henley, H.~R. Brashear, J.~Bogert, A.~Russu, L.~Janssens, I.~Tesseur, L.~Tritsmans \emph{et~al.}, ``Long-term safety and tolerability of atabecestat (jnj-54861911), an oral bace1 inhibitor, in early alzheimer’s disease spectrum patients: a randomized, double-blind, placebo-controlled study and a two-period extension study,'' \emph{Alzheimer's research \& therapy}, vol.~12, no.~1, p.~58, 2020.

\bibitem{yun2008t790m}
C.-H. Yun, K.~E. Mengwasser, A.~V. Toms, M.~S. Woo, H.~Greulich, K.-K. Wong, M.~Meyerson, and M.~J. Eck, ``The t790m mutation in egfr kinase causes drug resistance by increasing the affinity for atp,'' \emph{Proceedings of the National Academy of Sciences}, vol. 105, no.~6, pp. 2070--2075, 2008.

\end{thebibliography}
\end{document}